\documentclass[aps,prl,twocolumn,reprint %
 preprint, 
 amsmath,amssymb,
 aps, physrev,
]{revtex4-2}

\usepackage{graphicx}% Include figure files
\usepackage{dcolumn}% Align table columns on decimal point
\usepackage{bm}% bold math
\usepackage{tikz}

\begin{document}

\preprint{APS/123-QED}

%\title{\textbf{Gyrotropy-Induced Symmetry Breaking of Erosion Hot Spots on Antenna Limiters during High-Power Tokamak RF Operation: Mechanisms and Mitigation Strategies.} }%
\title{\textbf{Gyrotropy-Induced Symmetry Breaking in Tokamak RF Operation: Mechanisms and Mitigation Strategies.}}

\author{W. Tierens}
 \email{Contact author: tierenswv@ornl.gov}
\author{A. Kumar}%
\author{J. Lore}%
\author{K. Fujii}%
\affiliation{%
 Oak Ridge National Laboratory, 1 Bethel Valley Road,
Oak Ridge, TN 37830, USA
}%

\author{G. Urbanczyk}
\affiliation{
Institut Jean Lamour UMR 7198 CNRS-Universit\'e de Lorraine, 2 all\'ee Andr\'e Guinier, F-54011
Nancy, France
}%
\author{R. Diab}
\affiliation{%
 MIT Plasma Science and Fusion Center, Cambridge, Massachusetts 02139, USA
}
\author{The WEST Team}
\affiliation{See \url{http://west.cea.fr/WESTteam}}
%
%\author{R. Bilato}
%\affiliation{%
% Max Planck Institute of Plasma Physics
%}%


\date{\today}% It is always \today, today,

%PRL short abstract
\begin{abstract}
Ion Cyclotron Range of Frequencies (ICRF) heating is essential for creating plasma in next-generation fusion devices. ICRF antennas often produce hot spots, reducing reliability, material survivability, and overall plasma performance. We show that toroidal and poloidal asymmetries in these hot spots arise intrinsically from the wave physics' gyrotropy, and that they can be compensated for either by controlling the poloidal phasing or by modifying the limiter shapes, reducing total erosion by a factor $\sim$2 compared to state-of-the-art designs.\footnote{Notice: This manuscript has been authored by UT-Battelle, LLC, under contract DE-AC05-00OR22725 with the US Department of Energy (DOE). The US government retains and the publisher, by accepting the article for publication, acknowledges that the US government retains a nonexclusive, paid-up, irrevocable, worldwide license to publish or reproduce the published form of this manuscript, or allow others to do so, for US government purposes. DOE will provide public access to these results of federally sponsored research in accordance with the DOE Public Access Plan (\url{https://www.energy.gov/doe-public-access-plan}).}
\end{abstract}
\maketitle

%\tableofcontents

Ion Cyclotron Range of Frequencies (ICRF) is a leading technique for plasma heating and current drive in magnetic confinement fusion experiments, able to bring the plasma to reactor-relevant conditions \cite{garcia2024stable, Creely_2020}. Other applications include plasma-based space propulsion systems \cite{takahashi2022thrust, Arefiev_2004} and the study of plasma-material interactions \cite{rapp2019latest}. In 2022, ICRF alone contributed 20\% to the record 59MJ fusion energy produced during the last deuterium-tritium campaign at JET \cite{Maslov_2023}. In present-day fusion devices, ICRF performance depends critically on minimizing impurity generation and damage to plasma-facing components through optimized antenna design and operation.  Careful control of the antenna phasing and power ratio \cite{bobkov2019impact,diab2025mitigation} has proven effective in reducing material sputtering and impurity sources, making ICRF compatible even with the high-Z environment \cite{bobkov2016making} of most planned fusion devices \cite{barabaschi2025iter}. However, emerging experimental \cite{Colas_2022} and modeling results \cite{kumar2024integrated, tierens2024radiofrequency} indicate that these mitigation strategies may not fully capture certain asymmetries in the wave physics, which can cause uneven heat loads on the limiters, threaten material survivability and overall plasma performance.

\begin{figure}
    \centering
    \includegraphics[width=1.0\linewidth]{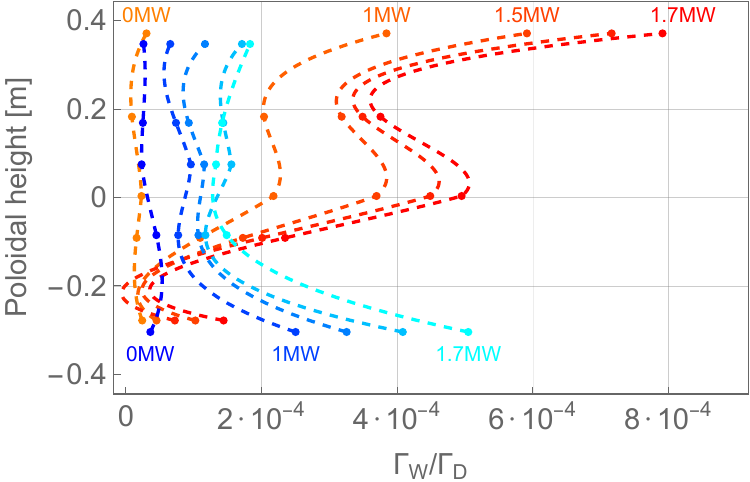}
    \caption{Sputtering yields during a power scan from 0MW through 1.7MW on the WEST antenna limiters (blue through cyan: left limiter, orange through red: right limiter), computed from the Tungsten ($\Gamma_W$) and Deuterium ($\Gamma_D$) fluxes observed using visible spectroscopy. The sputtering is both horizontally and vertically asymmetric. This work does not give a complete account of all sources of asymmetry in the experiment, instead we show that the wave physics itself still breaks both horizontal and vertical symmetry even if the antenna and the plasma are perfectly symmetric.%Both vertical asymmetry (neither the blue nor the red curves are vertically symmetric) and horizontal asymmetry (the red and blue curves have different height-dependence) are readily apparent.
    }
    \label{fig:1}
\end{figure}

Recent spectroscopic measurements of tungsten (W) and deuterium (D) fluxes on the WEST tokamak, shown in Figure~\ref{fig:1}, show highly asymmetric erosion patterns. These results are based on W and D line emission signals from chords viewing the antenna limiters during the ICRF-heated discharges \#61683 - \# 61689.  Similar asymmetric erosion  has also been observed in other fusion devices (e.g. AUG \cite{bobkov2009operation}, EAST \cite{zhang2022first} and NSTX  \cite{swain2003loading}), demonstrating the universal nature of this behavior in ICRF-heated magnetic fusion experiments. These observations are also consistent with recent numerical modeling \cite{tierens2024radiofrequency,kumar2024integrated, kumar_rfppc_2025, ragonaRFPPC2025}, which likewise predicts strongly asymmetric erosion hotspots at the antenna limiters. In this Letter, we show that an unavoidable asymmetry is inherent in the gyrotropy of the RF wave physics, persisiting even when other causes such as the antenna geometry, plasma shape, antenna-plasma misalignment, asymmetric excitation of the antenna ports, and the preferential poloidal direction of motion of convective cells, are take into account. This intrinsic effect fundamentally limits the effectiveness of symmetric or flux-surface-aligned limiters. We therefore  propose that non-aligned limiters, compensating for the gyrotropy of the wave physics, may outperform the current state of the art.

\begin{figure}
    \centering
  \includegraphics[width=\linewidth]{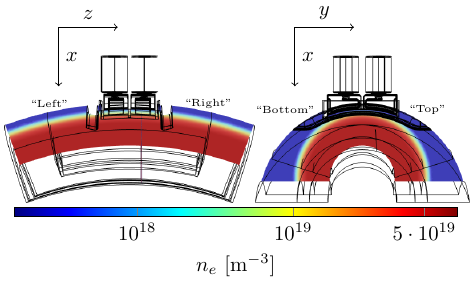}
    \caption{The symmetric plasma density in front of the symmetric antenna. Left: toroidal cross section, right: poloidal cross section.}
    \label{fig:neAndYeff}
\end{figure}

\begin{figure}
    \centering
    \includegraphics[width=0.47\linewidth]{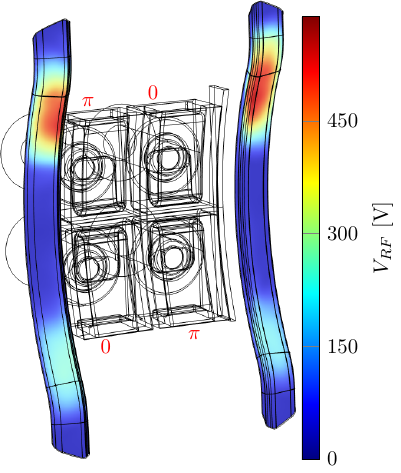}
    \includegraphics[width=0.51\linewidth]{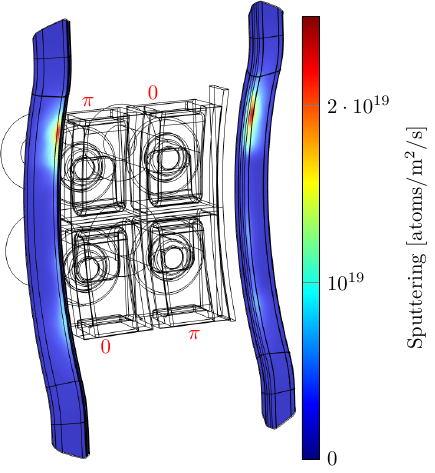}
    \caption{Left: Numerically predicted RF potential at 1MW coupled power, and (right) corresponding sputtering yield in a completely symmetric WEST-like antenna. Despite the symmetry, hot spots preferentially appear at the top. The total (integrated) sputtering yield is $5.2\cdot 10^{17}$ atoms per second. The phase of the incident wave in each of the four coaxial ports is indicated in red.}
    \label{fig:2}
\end{figure}

Numerically, it is possible to disentangle the various sources of asymmetry. We have constructed a fully symmetric WEST-like finite element antenna model based on the one used in reference~\cite{tierens2024radiofrequency}, removing all asymmetric elements, and ensuring the bottom half is the mirror image of the top half. We then consider a perfectly symmetric plasma using the density from figure \ref{fig:neAndYeff}, and a purely toroidal magnetic field (no tilt angle). We include sheath rectification physics using the dielectric layer sheath model \cite{beers2021rf}, which gives us both an RF potential $V_{RF}$ and a DC potential $V_{DC}$ on the limiters. Our antenna has four coaxial ports, in each of which we can specify the amplitudes and phases of the incident wave. We excite the antenna in a so-called dipole phasing \cite{urbanczyk2021rf} (wave phases $0,\pi,\pi,0$, red in figure \ref{fig:2}), with the same power on the top and bottom ports. Even in this maximally symmetric scenario, numerical results still predict up-down asymmetries in the rectified potential and the sputtering yield at the antenna limiters, as shown in figure
\ref{fig:2}. We computed the sputtering yield from the potential using a WEST-typical sputtering curve as in reference~\cite{tierens2024radiofrequency}. This up-down asymmetric erosion is robust to small changes in the vertical location of the plasma and small changes in the magnetic field tilt angle. Only by turning the confining magnetic field all the way around (reversing the direction of the currents in the toroidal field coils), which is equivalent to mirroring the situation since $B$ is a pseudovector, do the hot spots end up on the bottom instead.

%\begin{figure}
%    \centering
%    \includegraphics[width=0.75\linewidth]{yeff.pdf}
%    \caption{Sputtering curve $Y_{eff}(V_{DC})$ for WEST-like plasmas (Deuterium with 4\% H, 0.05\% B, 0.05\% C,
%and 0.001\% W), used to compute the sputtering yields in figures %\ref{fig:2} and \ref{fig:5}.}
 %   \label{fig:yeff}
%\end{figure}

This remaining asymmetry is actually inherent in the wave physics itself. Recall the electromagnetic frequency-domain wave equation in Cartesian coordinates
\begin{align}
    \nabla\times\nabla\times \boldsymbol{E}-k_0^2 \varepsilon \boldsymbol{E}=0 \label{we}
\end{align}
where $k_0=\omega/c$ is the vacuum wavenumber, $\omega$ is the wave (angular) frequency, $c$ is the speed of light, and the cold plasma dielectric tensor (in the absence of a magnetic field angle, with the confining magnetic field along $z$) is
\begin{align}
    \varepsilon=I-\sum_s \left[\begin{matrix}
        \frac{\omega_s^2}{\omega^2-\Omega^2} & \frac{\Omega_s}{-i\omega}\frac{\omega_s^2}{\omega^2-\Omega_s^2} & 0 \\ \frac{\Omega_s}{i\omega}\frac{\omega_s^2}{\omega^2-\Omega_s^2} & \frac{\omega_s^2}{\omega^2-\Omega_s^2} & 0 \\ 0 & 0 & \frac{\omega_s^2}{\omega^2}
    \end{matrix}\right] \label{dielectricTensor}
\end{align}
where $I$ is the unit matrix, and $\omega_s,\Omega_s$ are the plasma frequency and the cyclotron frequency of particle species $s$, respectively. The presence of off-diagonal elements in (\ref{dielectricTensor}) is referred to as gyrotropy, which vanishes in the highly magnetized limit \cite{becache2017stable}.

We can see that if $\varepsilon_{12}=-\varepsilon_{21}=0$ (the non-gyrotropic case), the wave equation (\ref{we}) possesses a mirror symmetry: if $\varepsilon$ is vertically symmetric ($\varepsilon(x,y)=\varepsilon(x,-y)$) and
\begin{align}
    \boldsymbol{E}=[E_x(x,y),E_y(x,y),E_z(x,y)]^T \exp(i n_z z \omega/c) \nonumber
\end{align}
is a solution of (\ref{we}), then so is its vertical mirror image
\begin{align}
    \boldsymbol{E}=[E_x(x,-y),-E_y(x,-y),E_z(x,-y)]^T \exp(i n_z z \omega/c) \nonumber
\end{align}
Thus, the absence of gyrotropy suffices to restore the vertical symmetry in our scenario.

\begin{figure}
    \centering
    \includegraphics[width=0.49\linewidth]{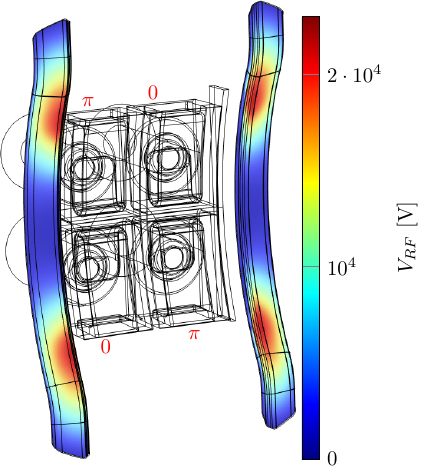}
    \includegraphics[width=0.49\linewidth]{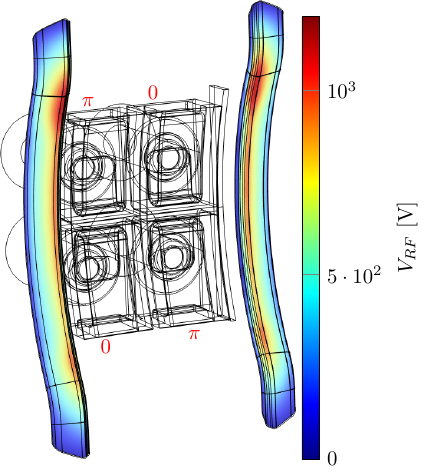}
    \caption{Eq. (\ref{ODE}) suggests two ways of restoring the vertical symmetry absent in figure \ref{fig:2}. Left: by artificially removing the gyrotropy. Right: by removing the gradients (constant density $n_e=10^{18}$m$^{-3}$ and magnetic field strength $B=3.2$T, though the antenna is still in vacuum). Only removing the gyrotropy succeeds in completely restoring the symmetry.}
    \label{fig:3}
\end{figure}

A deeper insight can be obtained by considering the two wave modes supported in the cold magnetized plasma: the fast wave and the slow wave \cite{stix1992waves}. For the fast wave, we have $E_z\approx0$. Taking $\boldsymbol{E}=[E_x(x),E_y(x),0]^T\exp(i (n_x x+n_y y + n_z z)\omega/c)$, and $\varepsilon=\varepsilon(x)$, we may derive an approximate expression for $E_y(x)$, proceeding along the lines of \cite{swain2003loading}:
\begin{align}
    \frac{A}{k_0^2}\frac{d^2 E_y}{dx^2} + \frac{B}{k_0}\frac{d E_y}{dx}  + C E_y = 0 \label{ODE}
\end{align}
with
\begin{align}
    A&=\frac{n_z^2-\varepsilon_{11}(x)}{n_y^2+n_z^2-\varepsilon_{11}(x)} \nonumber \\
    B&=-n_y^2\frac{\frac{d}{dx}\varepsilon_{11}(x)}{k_0(n_y^2+n_z^2-\varepsilon_{11}(x))^2} \nonumber \\
    C&=\frac{-n_y/(i k_0)}{n_y^2+n_z^2-\varepsilon_{11}(x)}\left(\frac{d}{dx}\varepsilon_{21}(x)+\frac{\varepsilon_{21}(x) \frac{d}{dx}\varepsilon_{11}(x)}{n_y^2+n_z^2-\varepsilon_{11}(x)}\right) \nonumber \\
    &-(n_z^2-\varepsilon_{11}(x))-\frac{\varepsilon_{21}(x)^2}{n_y^2+n_z^2-\varepsilon_{11}(x)} \nonumber
\end{align}
All coefficients in (\ref{ODE}) are symmetric in $n_y$, with the exception of $C$, which is symmetric in $n_y$ only if the plasma is non-gyrotropic ($\varepsilon_{21}=0$), or if there are no gradients in the $x$ (radial) direction. Numerical experiments in figure \ref{fig:3} show that only the first option (no gyrotropy) completely restores the symmetry. In fact, the mere presence of a plasma-vacuum interface (the antenna is kept in vacuum in all calculations in this work, hence there is a discontinuous plasma-vacuum interface) suffices to break the vertical symmetry, even if the plasma is otherwise uniform, which can be seen from earlier studies of the scattering problem on this interface \cite{tierens2021slab}.

Any parasitically excited slow waves, which may play an important role in the sheath excitation physics \cite{colas2025numerical,myra2021tutorial}, also have this asymmetry. The electric field of the slow wave can be approximated using a potential $\phi$ \cite{bellan2008fundamentals}. The electric field is $\boldsymbol{E}=-\nabla(\phi(x)\exp(i (n_y y + n_z z)\omega/c))$, and it obeys obeys $\nabla\cdot(\varepsilon \boldsymbol{E}) = 0$:
\begin{align}
    k_0^2 \phi (x) \left(n_y^2 \varepsilon_{11}(x)+n_z^2 \varepsilon_{33}(x)\right)&=\left(\frac{d}{dx}\varepsilon_{11}(x)\right)\left(\frac{d}{dx} \phi (x)\right) \nonumber\\
    &+ \varepsilon_{11}(x) \frac{d^2}{dx^2}\phi (x) \nonumber\\
    &+ i k_0 n_y  \phi (x) \frac{d}{dx}\varepsilon_{12}(x)
\end{align}
Once again, all coefficients are symmetric in $n_y$ provided that either $\varepsilon_{12}(x)=0$ (no gyrotropy) or $\frac{d}{dx}\varepsilon_{12}(x)=0$ (no gradients).

Returning to the horizontal symmetry, which is also absent in experiment (figure \ref{fig:1}), we consider under what circumstances the wave equation (\ref{we}) possesses a horizontal mirror symmetry. With the dielectric tensor (\ref{dielectricTensor}) a horizontal mirror symmetry exists even in gyrotropic plasma, but this symmetry is broken by the introduction of a nonzero magnetic field tilt angle. Thus, in the experimentally relevant scenario where the plasma is both gyrotropic and the confining magnetic field is not purely toroidal, we should expect neither horizontal nor vertical symmetry.

\begin{figure}
    \centering
    \includegraphics[width=1.0\linewidth]{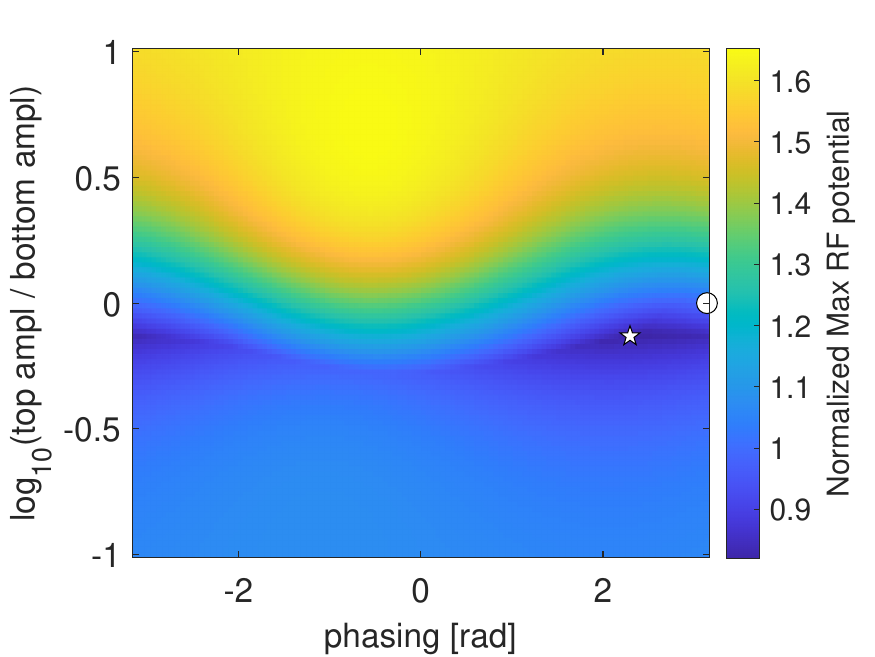}
    \caption{One way to compensate for the hot spots in figure \ref{fig:2} is to change the \emph{poloidal} phasing and power balance. The operating regime on WEST is $(\pi,0)$ (circle) on this plot: same amplitude but opposite phase on the top and bottom ports. The optimum (star) has slightly less power on the top port and a phase difference slightly below $\pi$. The RF potentials in this plot are normalized to the value at $(\pi,0)$.}
    \label{fig:phasingscan}
\end{figure}

Finally, let us consider some possible mitigation strategies. Hot spots like those in figure \ref{fig:2} can be reduced by controlling the phasing and power balance, as is often done to great effect in the toroidal direction \cite{bobkov2019impact}. Here, we have to do it in the poloidal direction. This is not possible on current devices, though certain benefits of having this capability were pointed out as early as the 1980s \cite{ram1984coupling}, and the future ITER ICRF antenna and the under construction WEST Travelling Wave Antenna \cite{Ragona_2022} will be able to do it. Figure \ref{fig:phasingscan} suggests that the peak RF potential can be reduced by $\sim 20\%$ following this approach.

\begin{figure*}
    \centering
    \includegraphics[width=0.3\linewidth]{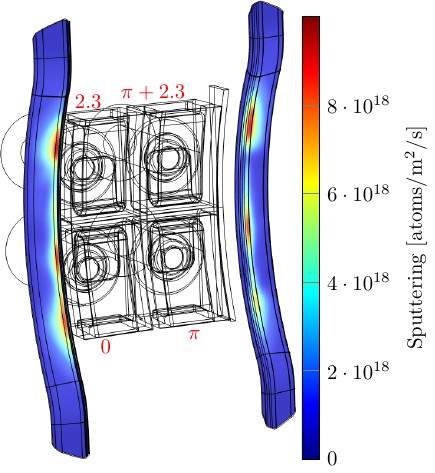}
    \includegraphics[width=0.3\linewidth]{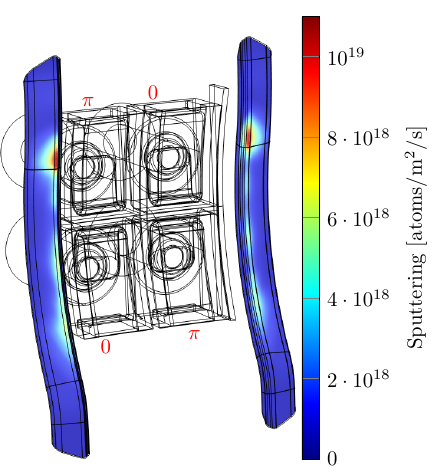}
\begin{tikzpicture}[scale=0.75, every node/.style={scale=0.75}]
  \node[anchor=south west, inner sep=0] (img) at (0,0) {\includegraphics[width=0.25\linewidth]{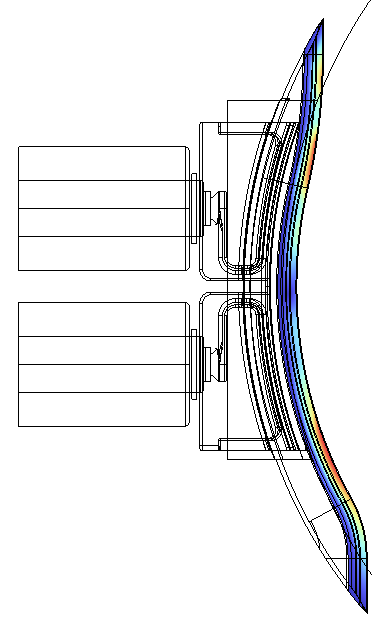}};
  \draw[blue, thick] (2.7,0.7) -- (4.5,0.7);
  \draw[blue, thick] (2.7,0.7) -- (4.5,1.6);
  \draw[blue, thick] (-3,6.57) -- (4.5,6.57);
  \draw[blue, thick] (-3,6.57) -- (4.5,4.8);
  \draw[blue, thick] (3.0,0.7) arc (0:24:0.3)  node[anchor=north]{24 deg, 0.15m};
  \draw[blue, thick] (-2.5,6.57) arc (0:-13:0.5) node[anchor=south]{13 deg, 1m};
\end{tikzpicture}
    \caption{Predicted sputtering on (left) the standard limiter with the optimal phasing from figure \ref{fig:phasingscan} (incident wave phase indicated in red for all four ports) and (middle) on a modified asymmetric limiter with standard phasing, whose geometry is shown on the right. The total (integrated) sputtering yield is $4.2\cdot 10^{17}$ (left) resp. $2.8\cdot 10^{17}$ (right) atoms per second.}
    \label{fig:5}
\end{figure*}

Another option is to modify the limiter shape. In light of the asymmetries inherent in the wave physics, we consider if an asymmetric limiter might outperform a symmetric one. We have parameterized the limiter shape in our model using three circular arcs, so we can retract it further away from the plasma at the location of the hot spots. Figure \ref{fig:5} (right) shows one possible modified limiter shape.
This figure also shows the predicted sputtering for both proposed modifications (poloidal phasing, modified limiter shape). Optimal poloidal phasing gives a $\sim$60\% reduction in peak sputtering and a $\sim$20\% reduction in total sputtering w.r.t. the reference of figure \ref{fig:2}. The modified limiter shape gives a factor $\sim$ 3 reduction in peak sputtering and a factor $\sim$ 2 reduction in total sputtering. Both cases show a more evenly distributed erosion pattern, and hence better material survivability.% This strongly suggests the possibility of a substantial further reduction in ICRF-specific impurity sputtering through the use of carefully designed antenna limiters, which should take the asymmetries inherent in the physics into account.

\iffalse
\begin{figure}
    \centering
\begin{tikzpicture}%[scale=0.75, every node/.style={scale=0.75}]
  \node[anchor=south west, inner sep=0] (img) at (0,0) {\includegraphics[width=0.5\linewidth]{nsl2.png}};
  \draw[blue, thick] (2.7,0.7) -- (4.5,0.7);
  \draw[blue, thick] (2.7,0.7) -- (4.5,1.6);
  \draw[blue, thick] (-3,6.57) -- (4.5,6.57);
  \draw[blue, thick] (-3,6.57) -- (4.5,4.8);
  \draw[blue, thick] (3.0,0.7) arc (0:24:0.3)  node[anchor=north]{24 deg, 0.15m};
  \draw[blue, thick] (-2.5,6.57) arc (0:-13:0.5) node[anchor=south]{13 deg, 1m};
\end{tikzpicture}
    \caption{The limiter shape in our model is parameterized using three circular arcs, allowing us to test a modified limiter shape where the top, where the hot spots occur, is further away from the plasma. Note that the antenna box remains entirely in the limiter shadow.}
    \label{fig:4}
\end{figure}
\fi

The predicted effectiveness of this apparently minor limiter shape change can be understood in terms of two factors: first, the edge density gradient is very steep, so even a small displacement gives rise to a large difference in the incident ion flux. Second, we displace precisely the location of the hot spot, which accounted for most of the sputtering to begin with. This allows us to reduce sputtering while keeping the overall coupled power constant, which is difficult to achieve when one retracts the entire antenna instead (this is possible on WEST). In machines where the ICRF actuators are carefully aligned with the first wall (e.g. ITER), such limiter deformation is not possible. On the other hand, in machines like WEST, C-Mod, or SPARC, where there is no first wall but only objects with distinct limiters poking into the plasma, this deformation can be successfully used to mitigate sputtering.

It is now standard in ICRF operation to control the toroidal phasing and power balance, which minimizes the parallel electric field near the limiters and thereby the sheath rectification \cite{bobkov2019impact}, and which can also be used to mitigate convective cell-induced asymmetries in the particle flux to the ICRF antenna \cite{diab2025mitigation}.
Our results strongly indicate that a further decrease of the RF-specific impurity sputtering, beyond that already achieved by these measures, is achievable by poloidal modifications of the antenna, ranging from relatively simple poloidal phasing and power balance control, to redesign of the limiter shape. This reduction is possible even in scenarios where the plasma is otherwise well-aligned with the antenna. This provides a  pathway to mitigate asymmetric RF-induced erosion hotspots (critical for material survivability) while also reducing the net erosion flux by nearly 50\%, thereby ensuring robust ICRH antenna operation, extending material lifetimes, and contributing to sustained reactor-grade plasma performance.
%This provides a pathway to handle the asymmetric hotspots which are critical for material survivability, to reduce overall erosion flux, and to ensure robust ICRH antenna operations and overall reactor grade plasma performance.
%This also achieves a more even distribution of the heat loads on the antenna surfaces, improving material survivability.

This manuscript has been authored by UT-Battelle, LLC,
under Contract No. DE-AC05-00OR22725 with the U.S.
Department of Energy (DOE). This material is based upon
work supported by the U.S. Department of Energy, Office of Science, Office of Advanced Scientific Computing Research and Office of Fusion Energy Sciences, Scientific Discovery through Advanced Computing (SciDAC) program.

\nocite{*}

\bibliography{apssamp}% Produces the bibliography via BibTeX.

\end{document}